\documentclass[showpacs,aps,prb,superscriptaddress,floatfix]{revtex4}
\usepackage{graphicx}    

\begin{document}

\title{Quantum approach to nucleation times of kinetic Ising ferromagnets}

\author{M. D. Grynberg} 

\affiliation{Departamento de F\'{\i}sica, Universidad Nacional de La Plata,
(1900) La Plata, Argentina}

\author{R. B. Stinchcombe}

\affiliation{Rudolf Peierls Centre for Theoretical  Physics, 
University of Oxford,1 Keble Road, Oxford OX1 3NP, UK}

\begin{abstract}
Low temperature dynamics of Ising ferromagnets under finite
magnetic fields are studied in terms of quantum spin representations
of stochastic evolution operators. These are constructed for the 
Glauber dynamic as well as for a modification of this latter, introduced
by K. Park {\it et al.} in Phys. Rev. Lett. {\bf 92}, 015701 (2004).
In both cases the relaxation time after a field quench is evaluated 
both numerically and analytically using the spectrum gap of the 
corresponding operators. The numerical work employs standard
recursive techniques following a symmetrization of the evolution
operator accomplished by a non-unitary spin rotation. The analytical
approach uses low temperature limits to identify dominant terms 
in the eigenvalue problem. It is argued that the relaxation times 
already provide a measure of actual nucleation lifetimes under 
finite fields. The approach is applied to square, triangular and 
honeycomb lattices. 
\end{abstract}


\maketitle

\section{Introduction}

Nucleation phenomena are of basic importance in a wide range of 
metastable systems which typically involve the crossing of a free  
energy barrier that is large compared to thermal fluctuations.\cite{reviews}
Classical  examples of such situation are the formation of droplets in an 
undercooled gas or of crystals in an undercooled liquid, whereas 
numerous analogies can also be found in contexts as  diverse as, 
for instance, material science \cite{ms},  astrophysics \cite{astro} and 
quantum liquids.\cite{ql}  Owing to the initial state of these systems, 
generally produced by a rapid quench from a stable phase,  the decay
time before escaping  from metastability  may result extremely large at 
low temperatures.  A significant part of the theoretical understanding of 
these relaxation processes has been amply developed in the study of 
kinetic Ising ferromagnets as microscopic (lattice) models of nucleation. 
In this framework, the metastable phase can be prepared after 
equilibrating the system under an external magnetic field $h$ which is
then suddenly reversed. The system therefore evolves towards the full  
minimization of its free energy via the formation of droplets or small
clusters of spins aligned with the new field direction. These droplets start 
growing with very small rates until at least one of them exceeds a critical
size, i.e. a saddle point configuration or a local maximum in the free 
energy landscape,  thus triggering a rapid magnetization change in the 
whole system. This stems in part from the competition between
the energy  gained by aligning spins with the field and the interface
energy created in reorienting previously parallel spins, thus escape from
metastability essentially occurs when the cost of the latter is outweighted
by the gain of the former.

Several analytical studies have been addressed  to elucidate the 
dynamical aspects of these processes in the low temperature limit 
\cite{Neves, Scoppola, Kotecky, Olivieri,Martinelli} while more recently,
the actual  evaluation of average nucleation lifetimes has been studied 
combining a  range of numerical and analytical efforts. \cite{Buendia, 
Park, Brendel, Shneidman, Novotny, Park2, Bovier} 
As a further step in this direction, in this work we 
discuss an alternative low temperature procedure (both numerical 
and analytical), to estimate the 
relaxation time $\tau\,$ of Ising ferromagnets evolving through detailed 
balance stochastic rules.\cite{Kampen} Specifically, we consider both the 
usual Glauber dynamic \cite{Glauber} along with a  seemingly minor 
modification of the latter which however yields entirely different 
characteristics at large magnetic fields.\cite{Park}  In either case, 
we construct a quantum spin representation of the evolution operator 
whose spectrum gap ($\tau^{-1}\,$) provides a measure of  nucleation 
rates.  In line with the general grounds referred to above, the implicit  
assumption allowing for this identification is that the first passage 
time\cite{Kampen}  to create randomly a critical 
nucleus is much longer than the characteristic timescale involved in 
subsequent growth\cite{Beijeren}. Thus, the relaxation of the entire 
system can be expected to coincide with the inverse of the probability 
of escaping from the metastable well. A posteriori, our results will
lend further support to this view.

Another assumption that is usually made in homogeneous systems
 -and which is crucial for the feasibility of our numerical approach- 
is that multi spin-flip events as well as fusion 
between subcritical clusters, are vanishingly rare in the low-temperature
limit\cite{reviews,Neves}. This is supported by our analytic work.
Therefore, the relevant length scale over which the slow part 
of the dynamic takes place is of the order of a critical droplet size, the 
first one to nucleate. Although on one hand this prevent us from dealing 
with small field regimes, where the nucleus becomes macroscopic
in the limit $h \to 0\,$, on the other hand  this enables us to study
other $h$-regions using numerically accessible clusters,
so long as the nucleus can be contained in them.
This does not presuppose either a precise knowledge of the nucleus 
size and shape (sometimes a conceptual problem of its 
own\cite{Kotecky,Wang}), or of the most probable path during a 
nucleation  event\cite{Kotecky,Neves}, so in this regard our 
numerical and analytical procedures provide a complementary approach
to that of absorbing Markov chains\cite{Kampen} and other related 
techniques discussed in Refs.\,[\onlinecite{Buendia,Park,Novotny,Bovier}].

For two-dimensional lattices and low temperatures $T\,$, the average
nucleation time $\tau (h,T)\,$  we aim to evaluate has been rigorously 
shown to be  parameterizable as\cite{Neves}
\begin{equation}
\label{tau}
\tau = A (h)\,\exp [\,\beta \,\Gamma (h)\,]\,,
\end{equation}
thus, the temperature dependence enters solely in  the factor
$1/\beta = k_B T\,$ (hereafter the Boltzmann constant $k_B$
is set equal to one). Independently of the stochastic dynamic considered,
the exponential argument $\Gamma\,$ has been often associated with the 
energy barrier separating the saddle point from the metastable phase. 
In fact, for the Glauber dynamics the analysis of Ref.\,[\onlinecite{Neves}]
corroborates this issue. However,  the results 
of  Refs.\,[\onlinecite{Buendia,Park}]  clearly indicate that such 
interpretation  of $\Gamma\,$ is not always  correct, even though the 
geometry of the  critical droplets remains unaltered by the change of 
dynamic. In particular, for the modified Glauber (MG) dynamics considered 
there\cite{Rikvold}, and to be discussed in the following sections,  despite 
detailed balance the value of $\Gamma\,$ does not actually yield that 
energy barrier. Moreover, under strong magnetic fields  the nucleation 
process led  by the MG dynamics turns out to remain active 
(i.e. $\Gamma > 0\,$), whereas above a maximum field the standard 
Glauber dynamic just  exhibits a fast decay ($\Gamma \equiv  0\,$). 
As for the amplitudes $A$ of 
Eq.\,(\ref{tau}), as well as for those involved in the nucleation times of 
a variety of systems\cite{reviews,Shneidman}, they have been usually 
difficult to evaluate numerically given their subdominant contribution. 
However, the lack of a fair estimation of these prefactors may introduce
far reaching theoretical implications in the nucleation picture\cite{Lothe}. 
Recent efforts have been addressed to remedy this situation
in the context of Ising ferromagnets under finite fields\cite{Shneidman,
Park,Buendia,Bovier}. On departing from the low temperature regime 
assumed in (\ref{tau}),  these amplitudes have revealed a structure of 
narrow peaks\cite{Shneidman} which however rapidly collapses towards 
a piecewise constant function of $h$ in the limit $T \to 0\,$. Their 
actual values also turn out to be dynamic dependent\cite{Park,Buendia}.
Here,  we numerically estimate the $\Gamma\,$ and $A\,$ parameters
under finite fields using both the Glauber  and MG dynamics introduced
in Ref.\,[\onlinecite{Park}]. To check the reliability of our numerical 
operational approach we compare its results with those obtained in 
square lattices\cite{Neves,Park}, and then proceed further in honeycomb
and triangular lattices where no results are previously available. One can 
expect that beside the evolution details, the relaxation parameters will be
also affected  by the  lattice structure as it determines the geometry of the
critical droplets ultimately controlling the nucleation 
time\cite{Kotecky,Neves}. Apart from the square lattice, their size and 
shape are not known  a priori but neither needed in our procedure.
Our low temperature analytic work provides results for $\Gamma$
and $A$ parameters generally in good agreement with their
numerical estimates.

The layout of this work is organized as follows. In Sec. II we recast the 
master equation governing the probability distribution of these processes
in terms of a quantum spin analogy whose 'Hamiltonian' provide the
appropriate transition rates between the original Ising spin configurations.
By means of an ulterior non-unitary spin rotation,  this results in a
symmetric representation of the evolution operator.  This simplifies 
considerably the subsequent numerical analysis of Sec. III  in which the
spectrum gap  of this latter representation is  obtained via standard 
recursive  techniques\cite{Lanczos} in several situations. 
In Sec. IV we develop the analytical approach, in which low temperature 
limits are used to pick out dominant terms in the hierarchy of equations 
obtained by applying the quantum spin Hamiltonian to an appropriate 
metastable state. We end the paper with Sec. V which contains our 
conclusions along with some remarks on extensions of this work.

\section{Dynamics and Operators}

Let us then consider an Ising ferromagnet with uniform nearest neighbor
(NN) interactions $J > 0 $ between the spins $s = \pm 1\,$ of a regular 
$d$-dimensional lattice. Under an applied magnetic field $h$, 
taken positive from now on, the corresponding Hamiltonian reads
\begin{equation}
\label{Ising}
{\cal H}_I =  - J \sum_{\langle {\bf r,r'}\rangle} s_{\bf r}\,s_{\bf r'}\,-\,
h \sum_{\bf r} s_{\bf r}\,,
\end{equation}
where the first and second sum run respectively over all NN or bond 
pairs $\langle {\bf r,r'}\rangle$, and all spin locations $\bf r$ of the 
lattice. Since ${\cal H}_I$ actually defines a classical energy functional, 
the constituents spins do not have a natural dynamics, i.e. 
$\partial_t \,s_{\bf r} = [\,{\cal H}_I, s_{\bf r} ] \equiv 0\,$.
Consequently, a specific stochastic evolution must be prescribed so as to 
emulate the interactions between the spins and a heat bath, here modeling 
the fast degrees of freedom not included in the classical Hamiltonian 
${\cal H}_I$. As usual,  the underlying non-equilibrium dynamics is then
approximated by a discrete Markovian process and therefore 
described by a master equation. The latter governs entirely the time 
evolution of the probabilities $P(s,t)$ of finding the system in a certain
spin configuration $\vert s \rangle$ at time $t$.  If $W(s \to s')\,$
denotes the (time independent) rate or transition probability per unit time 
at which configuration $\vert s \rangle\,$ evolves
to $\vert s' \rangle\,$, the master equation just adopts the continuity form
\begin{equation}
\label{master}
\partial_t \,P(s,t) = \sum_{s'} \left[\: W(s'\to s)\, P(s',t)\,
- \, W(s \to s') \,P(s,t)\:\right]\,.
\end{equation}

Because in the context of Sec.\,I metastability is imposed by an external
field,  among the decay processes representable by (\ref{master}) 
we restrict our attention to those in which the total magnetization is not 
preserved\cite{Hohenberg}.  One of the most studied and physically 
grounded\cite{Martin} examples of this type is the Glauber 
dynamic\cite{Glauber}. Its transition rates involve Ising configurations
differing at most in the state of a spin at a given site $\bf r$. With the
aid of the local field variables,  which henceforth we define as
$\varphi_{\bf r} =  \frac{J}{T} \sum_{\langle {\bf r, r'} \rangle}
 s_{\bf r'}\,$, these rates can be written as
\begin{equation}
\label{rGlauber}
W_G\, ( s_{\bf r} \to - s_{\bf r} ) = 
 \big[ 1+e^{2 \,(\varphi_{\bf r} + H)\,s_{\bf r}}\big]^{-1}\,, 
\end{equation}
where $H = h/T\,$. The iteration of these rules eventually bring the 
system to the Gibbs distribution as they clearly satisfy detailed
balance in Eq.\,(\ref{master}),  that is 
$W (s \to s') \,e^{- {\cal H}_I \{s\}} = W (s' \to s) \,e^{- {\cal H}_I
\{s'\}}$. However, other single spin flip or Glauber 
type processes  can also be made consistent with these latter condition,
so the approach to equilibrium in these problems is not unique. 
As it was referred to above, a recent case of this situation was introduced
in  Ref.\,[\onlinecite{Park}] with the aim of clarifying earlier issues of 
metastable lifetimes. In the MG dynamics proposed there, 
the effects of the $J$-interactions and the field $h$  are factorized in the
transition rates. More specifically, these are given by\cite{factor}
\begin{equation}
\label{rMG}
W_{MG}\, ( s_{\bf r} \to - s_{\bf r} ) = 
\big[ 1+e^{2\,\varphi_{\bf r}\,s_{\bf r}}\big]^{-1}\,
\big[ 1+e^{2\,H\,s_{\bf r}}\big]^{-1}\,.
\end{equation}
Although  it can be easily checked that such rates also comply with 
detailed balance, it will turn out that each of the above 
dynamics behaves quite differently under strong field regimes. 

{\it a. Mean field excursus} --
Before constructing a more convenient representation for these  
processes, we pause briefly to consider this latter dynamic at a simple 
mean field level of description.  Despite of being quantitatively 
uncontrolled, on the other hand it is able to account for some relevant 
qualitative features actually occurring in the MG dynamics 
(see Secs. III and IV). 
Thus, in order to decouple the rather  involved hierarchy of equations
implicit in Eq.\,(\ref{master}) we simply approximate the local field 
variables  $\varphi_{\bf r}\,$ by their mean value 
$z \langle \,s \,\rangle J/T\,$,  in turn assumed to be homogeneous. 
Here, $\langle \,s \,\rangle$  denotes the average magnetization whereas
$z$ stands for the number of NN spins, i.e. the lattice coordination 
number. After inserting  the so approximated $W_{MG}\,$ rates  in 
Eq.\,(\ref{master}),  we readily obtain the magnetization dynamic in terms 
of a non linear differential equation which at finite fields and low 
temperature regimes reduces to
\begin{equation}
\label{MF}
\partial_t \langle \,s \,\rangle =  \frac{1 - \langle \,s \,\rangle}{1 + 
e^{- 2\,z \langle s \rangle J/T }}\,,\hskip 0.5cm    h  /T  \gg 1\,.
\end{equation}
Hence, for the region of our interest the relaxation dynamics comes out 
to be field independent in this scheme. Although this is not the actual 
case below a minimum  $ h $-value (see results of Secs. III and IV), yet 
the analysis of Eq.\,(\ref{MF}) pinpoints a genuine difference with respect 
to the Glauber dynamics. Notice that for this latter, if $h /J > z\,$,  in the
limit $T \to 0\,$ the master equation results totally decoupled by the 
Glauber rates (\ref{rGlauber}), just as if the spins were independent. 
Then, it follows that  $\partial_t \langle \,s \,\rangle = 
1 - \langle \,s \,\rangle \,$, and therefore the time scale of the Glauber 
problem is of the order an elementary step, namely $\tau = 1\,$ (see also
Sec. III). By contrast, in Eq.\,(6) the magnetization evolves initially with a 
much slower pace as its change is exponentially plunged by the initial 
metastable phase. In fact, the integration of the reciprocal of
Eq.\,(\ref{MF}) between $\langle \,s \,\rangle  = -1\,$ and a subsequent
magnetization $ \langle \,s \,\rangle  >-1\,$ involves large escape times.
More precisely, with the aid of the exponential-integral function 
${\rm Ei}\, ( \langle s \rangle)\,$ and its asymptotic
expansions\cite{Gradshteyn}, we obtain
\begin{equation}
\label{tau-MF}
\tau \sim \frac{T}{4 z J} \; e^{2 z J/T}\,,\hskip 0.5cm   T/J \ll 1\,.
\end{equation}
This is consistent with a value of $\Gamma = 2 z J\,$ in Eq.\,(\ref{tau})
which later on will be corroborated both numerically (Secs. III)
as well as analytically (Sec. IV). 
The corresponding amplitudes however result significantly 
underestimated 
by this mean field simplification which nevertheless, is already 
able to capture the metastability of MG dynamics, at least 
under strong field conditions.

{\it b. Quantum spin representations} -- 
We now build up an alternative representation of Eq.\,(\ref{master}) 
lending itself more readily for a numerical study in finite spin clusters 
which, as pointed out in Sec. I, can embody the nucleation time of much 
larger systems. Firstly, it is useful to recall the matrix elements of the 
evolution operator $\cal H$ associated to a generic Markovian process. 
In terms of transition rates, these elements are constructed 
as~\cite{Kampen}
\begin{eqnarray}
\label{non-diag}
\langle\,s'\,\vert\,{\cal H}\,\vert\,s\,
\rangle &=& -\,W(s\to s')
\hspace{0.4cm},\hspace{0.4cm} s \ne s'\,,\\
\label{conservation}
\langle\,s\,\vert\,{\cal H}\,\vert\,s\,\rangle &=&
\sum_{s'\ne s}\, W(s \to s')\,.
\end{eqnarray}
This permits to think the master equation in imaginary time
as a Schr\"odinger-like representation 
$\vert\,P(t)\,\rangle = e^{\,-{\cal H}\,t }\, \vert\,P(0)\,
\rangle\,$ in which the probability distribution 
$\vert\,P(t)\,\rangle =  \sum_s\, P(s,t)\,\vert\,s\,\rangle$
evolves according
to  the action of the evolution operator -here playing the role of the 
'Hamiltonian'-  on the initial state  $\vert\,P(0)\,\rangle\,$
 (in our case, a metastable Gibbs distribution opposing the new
field direction).
The specific form of $\cal H$ in either of the above dynamics 
can be straightforwardly found in terms of spin-$\frac{1}{2}$ Pauli 
matrices $\vec \sigma$  and interpreting the local field variables
involved in Eqs.\,(\ref{rGlauber}) and (\ref{rMG}) as
local field operators $\varphi_{\bf r}^z$
\begin{equation}
\varphi_{\bf r}^z = \frac{J}{T} \sum_{\langle {\bf r, r'} \rangle}\,
\sigma^z_{\bf r'}\,, 
\end{equation}
which just for convenience are taken diagonal, say  in the 
$\sigma^z$-representation. To connect two $z$-configurations of spins
differing in the state of site $\bf r$, and therefore to account for the off 
diagonal  elements (\ref{non-diag}),  we simply project the corresponding 
'rate operator' (set by $\varphi_{\bf r}^z$), in terms of the usual spin 
raising and lowering projectors $\sigma^+_{\bf r}, \sigma^-_{\bf r}\,$.
For example, using the Glauber rates (\ref{rGlauber}), the
operational counterpart of (\ref{non-diag}) will read
\begin{equation}
\label{off}
 \sum_{s,s',\; s \ne s'}\!\vert s' \rangle\, \langle s'
\vert {\cal H}_G \vert s \rangle\,\langle s \vert =
- \sum_{\bf r} \Big \{\, 
\sigma^+_{\bf r}\, \big [ 1+e^{-2(\varphi_{\bf r}^z + H)} \big]^{-1}  
+ \,\sigma^-_{\bf r}\, \big[ 1+e^{2(\varphi_{\bf r}^z + H)} \big]^{-1} 
\,\Big\}\,.
\end{equation}
Since $[ \varphi_{\bf r}^z, \sigma^{\pm}_{\bf r} ] = 0\,$, 
the above ordering of application is immaterial. On the other hand, 
conservation of probability requires the emergence of the diagonal 
elements (\ref{conservation}). They basically count the number of ways 
in which a given configuration $\vert s\rangle$ can evolve to different
states $\vert s'\rangle$ through a single spin flip. This can be properly 
tracked down by using the number operators $\hat n_{\bf r} = 
\sigma^+_{\bf r} \sigma^-_{\bf r}$ along with the weighting of each 
flip with its corresponding rate. For the Glauber case
the analog of Eq.\,(\ref{conservation}) then becomes
\begin{eqnarray}
\nonumber
\sum_s\,  \vert s \rangle \,\langle s \vert {\cal H}_G \vert s \rangle\,
\langle s \vert &=&  \sum_{\bf r} \Big \{\, 
\hat n_{\bf r}\, \big [ 1+e^{2(\varphi_{\bf r}^z + H)} \big]^{-1} + \;
\big[ 1 - \hat n_{\bf r} \big] \, \big[ 1+e^{-2(\varphi_{\bf r}^z + H)}
 \big]^{-1} \,\Big\}\\
\label{diagonal}
&=& \frac{1}{2} \,\sum_{\bf r} \,\big[ \, 1 -  \sigma^z_{\bf r}\,
\tanh\, (\varphi_{\bf r}^z + H)\,\big]\,,
\end{eqnarray} 
which together with Eq.\,(\ref{off}) completes the form of ${\cal H}_G$.
Certainly, the above reasoning is extensible to the MG dynamic as well.
The related evolution operator ${\cal H}_{MG}$ of this case thus finally 
turns out to be
\begin{eqnarray}
\label{MGlaub}
\nonumber
{\cal H}_{MG} = &-& \frac{1}{2}\,{\rm sech}\:H\,\sum_{\bf r} 
\Big [\,\sigma^+_{\bf r}\, e^H\, \big ( 1+e^{\,-2 \,\varphi_{\bf r}^z} 
\big)^{-1} \, + \, \sigma^-_{\bf r} \,e^{-H}\,\big ( 1 + e^{\,2 \,
\varphi_{\bf r}^z} \big)^{-1} \, \Big]\\
&+&\frac{1}{4} \,\sum_{\bf r} \,\big( \, 1 -  \sigma^z_{\bf r}\,
\tanh \varphi_{\bf r}^z \,\big)\,
\big( \, 1 -  \sigma^z_{\bf r}\,\tanh H \,\big)\,,
\end{eqnarray} 
which of course reduces\cite{factor} to ${\cal H}_G$ when $h=0$. Also, 
it can be easily verified that either of these operators remain invariant
under the spin inversion $\sigma^z \to -\sigma^z$ along with the field 
reversal $h \to -h$,\, as they should.  
Given the rather involved manner in which all spins are coupled  through 
the local field operators $\varphi_{\bf r}^z$,  exact analytic treatments 
of the spectrum of ${\cal H}_G$ or of ${\cal H}_{MG}$  under generic 
field and temperature conditions may seem unlikely, even in  
$d=1$.\cite{Hilhorst}  However, by exploiting low temperature limits
analytic procedures can be developed and applied, as shown in Sec. IV.
In addition, numerical progress can be made in fair system sizes by 
means of a suitable similarity transformation which we now discuss.

{\it c. Symmetric representations} -- 
As is known\cite{Kampen}, the detailed balance property of rates 
(\ref{rGlauber}) and (\ref{rMG}) warrants the existence of representations
in which  ${\cal H}_G$ and  ${\cal H}_{MG}$  are symmetric and thereby 
diagonalizable. Moreover, a {\it common} transformation for that purpose
can be found  for both dynamics. To this end, we rotate the corresponding 
operators  around the  $z$ spin direction using a site dependent pure 
imaginary angle
\begin{equation}
\label{angle}
\phi_{\bf r} = - i\, ( \,\varphi_{\bf r}  + H\,)\,,
\end{equation}
where the $\varphi$'s are the original scalar fields introduced in 
Eqs.(\ref{rGlauber}) and (\ref{rMG}). This rotation is produced by the 
non-unitary similarity transformation $U = e^{-i\,S}\,$ with 
$S = \frac{1}{2}\, \sum_{\bf r}  \phi_{\bf r} \, \sigma^z_{\bf r}\,$, 
which in turn  results in the direct product
\begin{equation}
\label{similarity}
U = \bigotimes_{\bf r}\, U_{\bf r}\,,\:\:\:\:\,
U_{\bf r}=  \left[\matrix{e^ {\,-\frac{1}{2} (\varphi_{\bf r} + H)} & 
 0 \cr 0  &  e^{\,\frac{1}{2}(\varphi_{\bf r} + H)} }\right]\,.
\end{equation}
While the diagonal terms of ${\cal H}_G$ and ${\cal H}_{MG}$ 
remain unaltered by $U$, it is straightforward to show that
\begin{equation}
U  \,\sigma^{\pm}_{\bf r}\, U^{-1} = 
e^{\mp (\varphi_{\bf r} + H)}\, \sigma^{\pm}_{\bf r}\,.
\end{equation}
From this latter relation, one can immediately verify that the
rotated Glauber operator ${\cal H}'_G = U {\cal H}_G U^{-1}\,$
can finally be casted in the symmetric form
\begin{equation}
\label{SGlaub}
{\cal H}'_G = \frac{1}{2} \,\sum_{\bf r} \,
\Big[ \, 1\,-\,\sigma^z_{\bf r}\,\tanh\, (\varphi_{\bf r}^z + H)\, - \,
\sigma^x_{\bf r}\,{\rm sech}\,(\varphi^z_r + H )\,\Big]\,,
\end{equation} 
whereas the rotated version ${\cal H}'_{MG} = U {\cal H}'_{MG} U^{-1}\,$ 
of the MG dynamics is also symmetric and comes out to be
\begin{equation}
\label{SMGlaub}
{\cal H}'_{MG} = \frac{1}{4} \,\sum_{\bf r} \,
\Big[ \, \big(\,1 - \sigma^z_{\bf r}\,\tanh\, \varphi_{\bf r}^z \, 
\big)\, \big(\,1 - \sigma^z_{\bf r}\,\tanh\,H \, \big) - \,
\sigma^x_{\bf r}\,{\rm sech}\,\varphi^z_r\:\, {\rm sech}\,H\,\Big]\,.
\end{equation} 

The formal analogy with the Schr\"odinger picture referred to above
now becomes more transparent, as all solutions of the master equation
are necessarily obtained as superpositions of eigenstates 
$\vert \psi_{\lambda}\rangle$ with {\it real} eigenvalues (or 'energies')
$\lambda \ge 0\,$ of Hermitian 'Hamiltonians'. In particular, the ground
states $\vert \psi_0 \rangle$ of both ${\cal H}'_G$ and  ${\cal H}'_{MG}$ 
coincide and are closely related  to the equilibrium Gibbs distribution. 
This is because $\langle\, \psi^{^{^{\!\!\!\!\sim}}}\,\vert  \equiv \sum_s
\, \langle\, s \,\vert$  is the  {\it left} steady state of the original 
stochastic operators (notice that their columns add up to zero),
and therefore $\langle \psi_0 \vert = \langle  \psi^{^{^{\!\!\!\!\sim}}} 
\vert U^{-1} =  \sum_s\, \langle\, s \,\vert e^{- \frac{1}{2}\,\beta \,
{\cal H}_I \{s\}}$, modulo a normalization factor $\sqrt Z$ involving the 
partition function of  the Ising energies (\ref{Ising}). It is thereby a simple 
matter to check that in our symmetric representation the dynamics of any 
classical quantity $\cal  A$ (which is already diagonal in the 
$\sigma^z$-representation, such as the  magnetization, the energy 
${\cal H}_I$ or any microscopic correlator), can be written as
\begin{equation}
\langle {\cal A}\, \rangle (t)  = \frac{1}{Z} \,\sum_{s} {\cal A}  \{s\} \,
e^{- \beta \, {\cal H}_I \{s\}} \,+\,\sum_{\lambda > 0} \, 
e^{-\lambda \,t}\,  \langle  \psi^{^{^{\!\!\!\!\sim}}} \vert U^{-1} {\cal A}\,
\vert\,\psi_{\lambda}\, \rangle\,\langle\,\psi_{\lambda}\, 
\vert \, U\,\vert\,P(0)\,\rangle\,.
\end{equation}
Thus, we see that the relaxation times discussed throughout Sec. I 
can be read off from the first excited level of the evolution operators 
constructed so far and whose numerical analysis we next turn 
to consider.

\section{Numerical Results}

The main advantage of the symmetric representations
(\ref{SGlaub}) and (\ref{SMGlaub}) is that their lower eigenmodes,
which are just the ones dominating the above non-equilibrium terms,
can be efficiently computed using recursion-type algorithms devised
for hermitian matrices, e.g. the Lanczos technique \cite{Lanczos}.
The latter is particularly appropriate to study system sizes capable
to accommodate critical droplets arising from not too small field regimes. 
Specifically, for a {\it square} lattice  the critical nucleus
is an  ${\cal L} \times ({\cal L} -1)$ rectangle of  overturned spins gathered
to a similar spin on one of its long sides of length
${\cal L} = [\,2J/h\,]\,$, where $[\,]$ denotes the integer part.\cite{Neves}
Hence,  in line with the general arguments of Sec. I
one could expect that for $h/J \agt 0.5\,$ the spectrum gap of at least
a  $5 \times 4\,$ spin system will suffice to yield actual values of 
nucleation times in the low temperature limit. 

Thus, starting from a random initial state but chosen orthogonal to the 
Gibbs-like distribution $\vert \psi_0 \rangle\,$ referred to above, we 
carried out the standard Lanczos procedure in such spin clusters using
periodic boundary conditions (hereafter, assumed throughout this 
Section). Let us first consider the Glauber operator (\ref{SGlaub}). 
In Fig.\,1 we show the results obtained
from its first excitation level $\lambda_1$,  i.e. above equilibrium, 
when varying the field $h/J \in (1,4)\,$ at low temperature regimes
$T/J \sim  0.2\,-\,0.4\,$. The nucleation time parametrization
conjectured by Eq.\,(\ref{tau}) here identified with $1/\lambda_1$,
is consistent with both the data collapse in the main panel as well as with
the linear  behavior evidenced in the inset.  In particular, the slopes of the 
latter detect three typical amplitude values which in turn are used as 
scaling factors in the main panel thus producing, as expected, 
the collapse of different curves. After a least square linear
fitting of our data,  the corresponding relaxation parameters within
the above temperature and field ranges are found to be
\begin{equation}
\label{hard-square}
\Gamma_G = \cases{ 16.(1)\, J - 6.(0)\, h\,, \;\;\, 1\alt \frac{h}{J}< 2\,,\cr 
                       8.(1)\,J - 2.(0)\,h\,, \;\;\;\;\, 
                        2 \le \frac{h}{J} \alt 4\,,}
\hskip 1cm
A_G = \cases{0.4(3)\,, \;\;\,1 \alt \frac{h}{J} < 2\,,\cr
                    1.9(4)\,, \;\;\, \frac{h}{J} = 2\,,\cr
                    1.3(3)\,, \;\;\,  2 < \frac{h}{J} \alt 4\,.}
\end{equation}
It should be mentioned that  {\it below} $T/J \sim 0.1$ and $h/J \sim 1\,$, 
the spectrum gap is gradually comparable to the numerical 
propagation of our roundoff errors, while the convergence of the 
Lanczos recursion becomes slow and erratic. Nonetheless, above those
regimes, where these problems do not show up, our results are already
in fair agreement with those of Ref.\,[\onlinecite{Park}] as well as with 
the low temperature analysis of Sec. IV. As conjectured earlier, size 
effects are negligible around this field region, at least judging 
from $3 \times 3\,, 4 \times 4\,$ and preliminary results in 
$6 \times 4$ spin arrays, all of which can enclose the critical droplets
depicted in Fig.\,1. In this regard, notice that the corresponding values of 
$\Gamma_G$ are consistent with both the surface tension and magnetic 
energy of such droplets, in turn recovering  the interpretation of 
$\Gamma$ as an energy barrier. Also by approaching the decoupling 
condition  $h/J = z\,$ from below, the low lying levels which were 
non-degenerate so far, closely approach one another, as they should, 
whereas $\Gamma_G \to 0$.

Bolstered by these consistency checks, we now turn our procedure to 
honeycomb and triangular lattices for which these nucleation parameters 
are not previously available (see also Sec. IV). Due to the roundoff 
limitations mentioned above, we restricted the computations respectively
to  $h/J \agt 0.5\,,\, T/J \agt 0.1$  and $h/J \agt 1.5\,,\, T/J \agt 0.3\,$.
For the first situation, Fig. 2 displays the results so obtained in an 
18-spin honeycomb cluster (schematized by its lower inset). 
These are in line with parametrization (\ref{tau}), and for which our 
numerical estimations yield
\begin{equation}
\label{hard-honeycomb}
\Gamma_G \sim \cases{ 14.(1)\,J - 10.(1)\, h\,, \;\;\; 0.5 
                            \alt \frac{h}{J} < 1\,,\cr
                       6.(0)\,J - 2.(0)\,h\,, \;\;\;\;\;\;\; 1 
                        \le \frac{h}{J} \alt 3\,, }
\hskip 1cm
A_G \sim \cases{0.1(6)\,, \;\;\; 0.5 \alt \frac{h}{J}< 1\,,\cr
                    2.2(1)\,, \;\;\; \frac{h}{J} = 1\,,\cr                    
                    1.3(3) \,,    \;\;\; 1 < \frac{h}{J}\alt 3\,.}
\end{equation}
Preliminary tests using 24-spin clusters showed no substantial
differences with these results. This conforms with the fact that 
within our accessible lower field bounds, the above $\Gamma_G\,$'s 
at most can entail a 5-spin nucleus (assuming the usual $\Gamma\!$- 
interpretation still holds). However, the shape of such nucleus can not be
inferred only from its surface tension $(14 J)$ as $z$ is not large 
enough.  In contrast, the results of the triangular lattice lend themselves 
more readily for this purpose, at least for the field range shown in Fig.\,3. 
After analyzing $4 \times 4$ and $5 \times 4$ triangular clusters, in this 
case further cusps in $\Gamma$ and amplitude discontinuities are 
detected, namely
\begin{equation}
\label{hard-triangular}
\Gamma_G  \sim \cases{ 32.(1)\,J - 10.(1)\, h\,, \;\;\, 1.5 
                 \alt \frac{h}{J} <  2\,,\cr 
                 20.(0)\,J - \;4.(0)\,h\,, \;\;\;\;\, 2 \le \frac{h}{J} < 4\,,\cr 
                 12.(0)\,J - \;2.(0)\, h\,, \;\;\;\;\,
                 4 \le \frac{h}{J} \alt 6\,, }
\hskip 1cm
A_G \sim \cases{0.7(0)\,, \;\;\;\,1.5 \alt \frac{h}{J} < 2\,,\cr
                    1.(0)\,, \;\;\;\;\;  \frac{h}{J} = 2\,,\cr
                    0.4(5)\,, \;\;\;\,  2 < \frac{h}{J} < 4\,,\cr 
                    1.4(2)\,,  \;\;\;  \frac{h}{J} = 4\,,\cr
                    1.2(2)\,,  \;\;\;  4 < \frac{h}{J} \alt 6\,.}
\end{equation}

Next, we consider the modified Glauber operator (\ref{SMGlaub}). In all 
studied situations, its numerical treatment comes out to be numerically 
more demanding,  i.e. spectrum gaps are even smaller than before, 
particularly below $h/J = z - 2\,$. So, we limit our computations to 
$h/J \agt 1 \,,\, T/J \agt  0.2$ for square lattices, 
$h/J \agt 0.3\,,\, T/J \agt 0.2$ (honeycomb) and, 
$h/J \agt 2 \,,\, T/J \agt 0.3$ (triangular). Despite these restrictions, the 
results of Fig.\,4 clearly support larger  values of $\Gamma (h)\,$
than those obtained for the Glauber dynamic.  Also,  the 
amplitude values turn out to be different as well as their regimes of validity.  
Specifically, for the square lattice (Fig.\,4a), we find
\begin{equation}
\label{soft-square}
\Gamma_{MG} \sim \cases{ 16.(1)\,J - 4.(0)\, h\,, \;\; 1< \frac{h}{J} < 2\,,\cr
                              8.(0)\,J\,,\;\;\;\;\;\;\;\;\;\;\;\;\;\;\;\;\;\;
                              \frac{h}{J} \ge 2\,, }
\hskip 1cm
A_{MG} \sim \cases{0.2(3)\,, \;\;\;\, 1 < \frac{h}{J} < 2\,,\cr
                  1.4(2)\,, \;\;\; \frac{h}{J} = 2\,,\cr 
                   1.(0)\,,    \;\;\;\;\; \frac{h}{J} > 2\,,}
\end{equation}
whereas for honeycomb (Fig.\,4b), and triangular (Fig.\,4c) systems the
respective parameters become
\begin{equation}
\label{soft-honeycomb}
\Gamma_{MG}
\sim \cases{ 14.(1)\,J - 8.(0)\,h\,, \;\;\; 0.3 \alt \frac{h}{J} < 1\,,\cr
         6.(0)\,J\,,\;\;\;\;\;\;\;\;\;\;\;\;\;\;\;\;\;\;\;  
\frac{h}{J}\ge 1\,, }
\hskip 1cm
A_{MG} \sim \cases{0.1(3)\,, \;\;\; 0.3 \alt \frac{h}{J} < 1\,,\cr
                    1.6(6)\,, \;\;\; \frac{h}{J} = 1\,,\cr                    
                    1.(0) \,,\;\;\;\;\;  \frac{h}{J} > 1\,,}
\end{equation}
and
%
\begin{equation}
\label{soft-triangular}
\Gamma_{MG}
 \sim \cases{ 20.(1)\,J - 2.(0)\,h\,, \;\;\;\, 2 \alt \frac{h}{J} < 4\,,\cr
         12.(0)\,J\,,\;\;\;\;\;\;\;\;\;\;\;\;\;\;\;\;\;\, \frac{h}{J} \ge 4\,, }
\hskip 1cm
A_{MG} \sim \cases{0.3(4)\,, \;\;\; 2 \alt \frac{h}{J} < 4\,,\cr
                    1.1(6)\,, \;\;\, \frac{h}{J} = 4\,,\cr                    
                    1.(0) \,,    \;\;\;\;\,  \frac{h}{J} > 4\,.}
\end{equation}
It is worth remarking that, as before, the square lattice parameters
are in reasonable agreement with those of Ref.\,[\onlinecite{Park}], 
which lend us further confidence in the identification of $1/\lambda_1$ 
with the nucleation time of the system.  Since in all studied cases 
$\Gamma_{MG}\, (h)  >  \Gamma_G\, (h)$, notice that the usual 
association of $\Gamma$ with an energy barrier no longer applies 
for this dynamic.\cite{Park,Buendia}  Also, these results  give evidence 
that the nucleation process persists at large times and fields, i.e. 
$\Gamma_{MG} = 2 J z > 0$,  as opposed to the Glauber picture where 
$\tau = 1$ beyond $h/J  = z\,$. Other tests using much larger fields 
suggest an identical behavior (see also Sec. IV),  always maintaining a
non-degenerate level $\lambda_1$.

Finally, we point out that errors throughout all  $A$'s  might be actually 
larger than those estimated above, given their subdominant contribution 
to $\tau\,$ in Eq.\,(\ref{tau}), specially within the smaller field regions.
This is reflected in the low sensitivity of the data collapse to amplitude 
changes slightly away from their error bands (arising only from upper 
insets).

\section{Low temperature analysis}

Here we develop a low temperature analysis, starting from the quantum 
formulation of Section II, which provides analytic results for 
the relaxation parameters $\Gamma$ and $A$.
The method employs the unsymmetrised Hamiltonian  
\begin{equation}
{\cal H} = - \sum_r \left[ (\sigma_r^+  -  P_r^-)\, R^+
\big(\sum_{r{\prime}} \sigma_{r{\prime}}^z\big)  +
(\sigma_r^-  -  P_r^+) \, R^- \, \big(\sum_{r{\prime}} 
\sigma_{r{\prime}}^z\big) \right]\,. \label{e:G}
\end{equation}
where $r{\prime}$ are the neighbours of $r$, and
$P_r^\pm \equiv {\frac{1}{2}}(1 \pm \sigma_r^z)$.
Low temperature versions of the appropriate rates $R^{\pm}(m)$ are
used.  Here, and throughout, $m$ is an integer corresponding to the 
"total spin" of the neighbours, and $\pm$ relates to flip up or down.
The unsymmetrised form of ${\cal H}$ is easier to work with because 
the low temperature forms of the rates there are nicely separated.

With $ \vert \,n\, \rangle$ the amplitude corresponding to a domain of $n$ 
up spins in  the eigenfunction for eigenvalue $s$, the eigenvalue problem
involves a hierarchy of equations relating $ \vert \, n \,\rangle$ to 
$\vert \,n \pm 1 \rangle$, each of which is of the following schematic form: 
\begin{equation}
\big[\,c_n \, R^+\,(\cdots) + d_n \, R^-\,(\cdots) - s\,\big] \, 
\vert \, n \rangle =  c_n \, R^+\,(\cdots) \, \vert \, n+1 \rangle + 
d_n \, R^-\,(\cdots)\, \vert \, n-1 \rangle. \label{e:Hier}
\end{equation}
Here the coefficients $c_n$ and  $d_n$ depend on geometric factors 
of the lattice such as its coordination number $z$. 
These equations are consistent with an equal amplitude 
eigenstate with eigenvalue $s = 0$. We want the next eigenvalue, 
i.e. the "gap" $s = (A \,e^{\beta \, \Gamma})^{-1}$. 

The relaxation from a metastable state, which we take to have all spins 
down (i.e. antiparallel to the field), is governed by the slow rates, 
especially the slow up-flip rates. Which rates are small depends on the 
field, so different field regimes have to be considered separately. 
The first equation  ($n = 0$) has $d_0 = 0$, 
i.e. the only terms come from $c_0 R^+\,(\cdots) = R^+\,(-z)$,
corresponding to nucleation of a single up spin, which state has amplitude 
$\vert 1 \rangle$.  The only case where $R^+ (-z)$ is not small at low 
temperatures is the Glauber dynamics case with $zJ < h$, where 
$R^+(m) \sim 1$, all $m$. The equations then give the gap of order 1, so  
\begin{equation}
\Gamma_G = 0 \;\; , \,\; h/J > z\,. \label{e:Reg0}
\end{equation}

In all other cases  $R^+\,(-z)$ is exponentially small in $\beta$ at low 
temperatures, and this results in non-zero $\Gamma$ for Glauber 
dynamics (G)  with $zJ > h$  and for modified Glauber dynamics (MG) 
at any $h$.  So we confine our attention hereafter to those cases, 
at very low temperatures. Then, $R^+\,(-z)$ is by far the smallest of the 
flip-up rates (since for any positive integer $l$, for MG dynamics 
$R^+\,(-l) \sim  e^{-2\beta lJ} \ll 1$, while in the Glauber case, 
 if $h < l\,J$ then $R^+\,(-l) \sim e^{-2\beta (l\,J - h)} \ll 1$). 
Consequently we may neglect terms involving  further factors of 
$R^+\,(-z)$, as occur, corresponding to further nucleations of isolated 
single-spin clusters, in the equations for $n > 0$. As a result, in 
Eq.\,(\ref {e:Hier}),  for amplitudes $\vert n \geq 2 \rangle$,
it is only necessary to consider "connected clusters" where all 
up-spins have at least one up-spin neighbour. For example the second 
equation ($n = 1$) has $c_1 R^+ (\cdots) = z R^+ \big(-(z - 2)\big)$ 
and  $d_1 R^- (\cdots) = R^- (-z)$ after neglecting the further 
nucleation terms involving  $R^+\,(-z)$.

While basic ideas and procedures are similar for G and MG dynamics, 
because of the different forms of their rates, the ordering of terms 
in the equations can be different in some regimes,  so we discuss the 
two cases separately, beginning with the Glauber case.

\subsection{Glauber dynamics}

{\it Case (a):} $d$-dimensional lattices with $zJ > h > (z - 2)J$. 
Here $R^+ (-z) \sim e^{-2\beta (zJ - h)} \ll 1$, 
and all other $R^+ (m) \sim 1$ while $R^- (-z) \sim 1$.

The resulting recursion type eigen-equations have (as always) the 
equal amplitude solution with $s = 0$, and (because the only small $R^+$ 
occurring is in the first equation) the next eigenvalue satisfies (in the low
temperature limit)  $(R^+ (-z) - s)(z + 1 - s) = R^+ (-z)$,  giving 
$s = z \, (z + 1)^{-1} e^{-2\beta (zJ - h)} = (A \,e^{\beta \,\Gamma})^{-1}$. 
So 
\begin{equation}
\Gamma_G = 2(zJ - h) \;\; , \;\; A_G = (z + 1)/z, \label{e:Reg1}
\end{equation}
for $zJ > h > (z - 2)J$ in any lattice. 
(This is consistent with the numerical predictions, and it includes the 
linear chain result $\Gamma_G = 4J - 2 h\,,  A_G = 3/2$).\cite{Hilhorst}

{\it Case (b):} $d > 1$-dimensional lattices  in the next regime, 
$(z - 2)J > h$. 

Now, as well as $R^+ (-z)$, a second flip-up rate, $R^+\,\big(-(z - 2)\big)$ 
becomes very small.  The geometry of the domains of $n$ and 
$n \pm 1$ sites determines the numbers $m$ in the rates $R^+(m)$ 
occurring in the equation for $ \vert n \rangle$. In particular the size, $q$, 
of the smallest ring of bonds on the lattice determines as $q - 2$ the 
number of successive equations in which the only up rate is 
$R^+ \big(-(z - 2)\big)$. The consequence for the gap is that s is 
proportional to $R^+\,(-z)\, [\,R^+\,\big(-(z - 2)\big)\,]^{q - 2}$, giving
the result: 
\begin{equation}
\Gamma_G = 2\,(z\,J - h) + 2\,(q - 2)\,\big [\,(z - 2)\,J - h\,\big]\,,
\label{e:Reg2}
\end{equation}
where $q = 4, 6, 3$ for square, honeycomb, and triangular lattices. 
These results agree with the numerical ones.

The determination of $A$'s is most easily carried out by considering the 
(first order, nonlinear) recurrence relations for 
$\mu _n = \vert n \rangle / \vert n - 1 \rangle$.
In the low temperature limit, ratios of small rates make the 
deciding $\mu _n$'s tend to 1, for the (q - 2) lowest n's, and to 0 
for the next one;  and in the limit the ratios of the  coefficients 
$c_n  ,  d_n$ in that range of n's determine the numerical factor 
in $(1 - \mu _1)$  and hence $A_G$. 
The (analytic, numerical) results are  $A_G =\,$ \big(3/8\,,  0.4(3)\big)\,, 
\big(1/6\,, 0.1(6)\big)\,, \big(1/3\,, 0.4(5)\big) for square, honeycomb, 
and triangular lattices,  respectively; the agreement is good except for 
the last one. 

The boundaries of the region are set by where new combinations of 
rates $R^{\pm}$ become limiting. The analytically determined ranges 
of validity of the results in this case {\it (b)} are $4J > h > 2J$ for
triangular, $2J > h > J$ for square, and $J > h > J/2$ for honeycomb 
lattice. 

{\it Case (c):} For the triangular lattice there is further regime 
($2J > h$) where yet another up-flip rate, namely
$R^+\, \big(-(z - 4)\big)\,$, becomes small.

Here we expect, in analogy to the argument and results given 
above for the previous regime that s will be proportional to 
$R^+\,(-z) \, \big[\,R^+\big(-(z - 2)\big)\,\big]^{q - 2}\, 
\big[\,R^+\,\big(-(z - 4)\big)\,\big]^Q\,$ with $Q$ an integer related 
to topological  features of the triangular lattice. This gives the form 
\begin{equation}
\Gamma_G = 2\,(\,z\,J - h\,) + 2\,(\,q - 2\,) \, \big[\,(z - 2)\,J - h\,\big] 
+ 2\,Q\, \big[\,(z - 4)J - h\,\big], \label{e:Reg3}
\end{equation}
(with $z = 6, q = 3$\,).  The numerical results are consistent with 
this with $Q = 3$.

\subsection{Modified Glauber dynamics}

Procedures for the MG dynamics are in principle similar. 
But now the primitive rate $R^+\,(-z) \sim  e^{-2\beta zJ}$ is always 
small,  in all regimes (even $h$ very large), so always
\begin{equation}
\Gamma_{MG} \neq 0. \label{e:reg0}
\end{equation}

For the following we use the notation  
$\epsilon = R^+\,(-z)$;  
$\Delta = R^+\, \big(-(z - 2)\big)$;
$\gamma = R^+\,\big(-(z - 4)\big)$;
$\delta = R^-\,(-z) \sim R^-\,\big(-(z - 2)\big)$;
$\mu _n = \vert \, n\, \rangle / \vert \, n - 1\rangle = (1 - \lambda_n)$. 
Then for any lattice the first two recurrence equations are 
\begin{eqnarray}
s &=& \epsilon \lambda_1\,,\\
z \,\Delta \lambda_2  &=& \delta\, \lambda_1/ \mu _1 + s\,.
\end{eqnarray} 
After these, the equations become lattice-dependent; 
e.g. for the triangular lattice the next ones are 
\begin{eqnarray}
2\, \gamma \,\lambda_3  &=& 2\, \delta\, \lambda_2/ \mu _2 + s\,,\\
3\, \gamma \, \lambda_4  &=& 3\, \delta\, \lambda_3/ \mu _3 + s\,,\\
4\, \gamma\, \lambda_5  &=& 3\, \delta \, \lambda_4/ \mu _4 + s\,,
\end{eqnarray}
etc.\, Details of the further reductions depend on the field regime.

{\it Case (a):} For $h > (z - 2)J$, the rates satisfy
$\delta \ll \Delta \ll \gamma$. 

Then for all the lattices the  equations give two possible forms of solution:
$\lambda_n \ll 1$, so $\mu _n \sim 1$ and $s = 0$ (ground state); 
or  $\mu_n \ll 1\,$, so $s = \epsilon \sim e^{-2\beta zJ}$.  Hence 
\begin{equation}
\Gamma_{MG} = 2zJ\;,\;  A_{MG} = 1. \label{e:reg1}
\end{equation}
This applies for any lattice, including the linear chain.

{\it Case (b):} In the next regime  $(z - 4)J < h < (z - 2)J$, possible in 
$d > 1$-dimensional lattices, the rate ordering is 
$\Delta \ll \delta \ll \gamma$. 

First consider the specific case of the triangular lattice. 
Again because  $\delta \ll \gamma$, for $s \neq 0$ the higher 
equations of the hierarchy give  $\lambda_n \sim 1, n=2,3,...$ But now  
$\Delta \ll \delta$, so the second equation of the hierarchy gives 
$\lambda_1 = z \Delta / \delta$  and it follows that
$s = z \epsilon \Delta / \delta$.  Hence 
\begin{equation}
A_{MG} = 1/z\; , \;\;\; \Gamma_{MG} = 2\,J\,z + 2\,J\,(z - 2) - 2\,h = 
20\,J - 2\,h, 
\label{e:tri}
\end{equation}
for the triangular lattice in this regime.

For the other lattices, a cycle of $(q - 2)$ successive equations 
(after the first equation) involve $\Delta / \delta$ and that is the 
origin of the general form
\begin{equation}
\Gamma_{MG} = 2\,z\,J + 2\,(q - 2)\, \big[\,(z - 2)\,J - h\,\big]. 
\label{e:reg2}
\end{equation}
The (analytic, numerical) results for $A_{MG}$ for
$(z - 2)J > h$, are \big(1/8, 0.2(3)\big)\,, \big(1/6, 0.1(3)\big)\,, 
\big(1/6, 0.3(4)\big) for square, honeycomb, and triangular lattices, 
respectively.

{\it Case (c):} On the boundary $ h = (z - 2)J$ between the last two 
regimes. Here the (analytic, numerical)  results for $A_{MG}$  are
 \big(11/8, 1.4(2)\big)\,, \big(11/6, 1.6(6)\big)\,, \big(7/6, 1.1(6)\big) 
for square, honeycomb, and triangular lattices, respectively.

\section{Conclusions}

The low temperature relaxational kinetics of Ising ferromagnets in 
a field has been treated for various lattices for both Glauber(G) and 
modified Glauber (MG) processes using a quantum representation 
of the evolution operators. The unsymmetrised version is convenient 
for the analytical work in the low temperature limit (Section IV) 
while the recursive numerical approach (Section III) requires the
symmetrised form produced by a special spin rotation.

The gross features produced by the two approaches agree completely. 
Those features include striking differences between the Glauber and 
modified Glauber cases: the latter case is always activated, even at 
high fields; in each of a sequence of field regimes, for both processes 
$\Gamma$ is found to be a linear function of $h$ and the amplitude 
$A$ is a constant, but  both $A$ and the form of $\Gamma$, and also 
the field regimes, differ between Glauber and modified Glauber cases.

While $A$ and the slope of $\Gamma (h)$ are discontinuous at regime 
boundaries, $\Gamma$ is continuous. This can be understood from the 
analytic discussion, where it is seen that the regimes are distinguished by 
which rates are most limiting, and in the low temperature limit the 
exponents in the limiting rates cross over at the regime boundaries, 
and those exponents determine $\Gamma$ but not $A$.

Both the numerical and analytic investigations make no use of the
shape of critical droplets nor of the most probable path towards
a nucleation event, and were carried out for square, triangular, and 
honeycomb lattices. The numerical approach indicates, and the analytic
one confirms, that $\Gamma$ is lattice-dependent (except in the 
non-activated high field regime of the Glauber case, where it is 
zero), and also $A$ is lattice-dependent except in the highest field regime
for the modified Glauber case. According to the analytic work, for the 
highest field activated regime the lattice-dependence of $\Gamma$ 
involves just the coordination number $z$ for both Glauber and modified 
Glauber cases, but in subsequent regimes other geometrical aspects 
of the lattice, e.g. the smallest ring size $q$, affect the value of $\Gamma$.
Similarly the lattice-dependence of $A$ is, as one moves down the field 
regimes, first through $z$, and thereafter involving further aspects of the
lattice. 

The quantitative agreement between the predictions of the numerical and
analytic approaches is very good for $\Gamma$, and for the regime 
ranges, and slightly less good for $A$, particularly for the smallest field 
regimes. This is as might be expected, since (i) at low temperatures 
a given (but bounded) numerical error in the gap evaluation will mostly 
propagate an error in $A$, rather than in $\Gamma$, as the latter carries 
an extra weight proportional to $\beta$; and (ii) as $h$ decreases the gap
becomes smaller, and so does the accuracy of the machine calculations. 
To understand more fully the comparisons between the numerical and 
analytic results it would be desirable to generalise the analytic work to 
finite low temperatures. 

The analytic work includes predictions for arbitrary lattices (e.g. for the 
regimes in which the results for $\Gamma , A$ depend only on $z$). 
It would be valuable to extend this, and to extend the numerical work to
other, especially three-dimensional, lattices. 
Further suggested extensions of the work are to other models (e.g. 
Potts models, where domain walls remain sharp) and to disordered 
cases: even a low concentration of weak bonds can make nucleation 
much faster.

\section*{Acknowledgments}

This work was supported by EPSRC under the Oxford Condensed 
Matter Theory Grants GR/R83712/01 and GR/M04426.
M.D.G. acknowledges financial help and kind hospitality of the 
Department of Theoretical Physics, Oxford, UK, where the later 
stages of this work were carried out. Support of CONICET, Argentina,
(PEI 1163 and PIP 2763),  is acknowledged.


\newpage

\begin{figure}
\vskip 1.5cm
\hskip -0.5cm
{\centering \resizebox*{4.7in}{!}{\includegraphics*{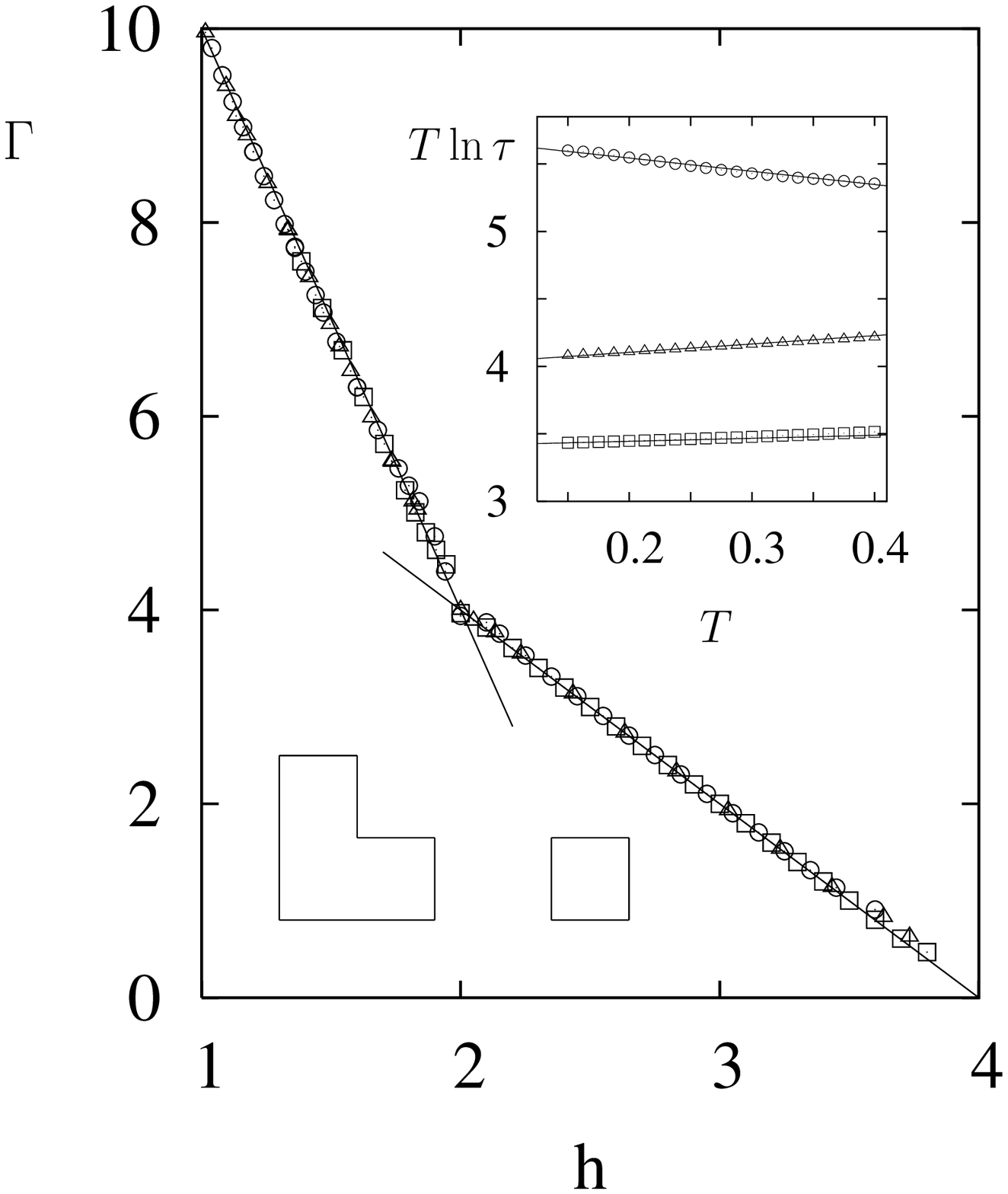}}}
\vskip -2cm
\caption{Low temperature estimation of the relaxation parameters
$\Gamma \equiv T \ln (\tau/A)\,$ in square lattices for $T = 0.2\,$ 
(squares), 0.3  (triangles) and 0.4 (circles) resulting from the Lanczos
diagonalization of Eq.\,(\ref{SGlaub}) in small clusters (up to $6 \times 4\,$ 
spins).  Solid lines denote the estimations of Eq.\,(\ref{hard-square}).
The data  collapse was attained upon using the amplitudes $A = e^s$ 
derived from the slopes $s$ of the upper inset. From top to bottom they 
refer respectively to $h = 1.7,\, 2\,$ and $2.3\,$, characterizing typical 
regimes of (\ref{SGlaub}) . The lower lines sketches the shape of the 
critical droplets for $1< h < 2$ (three spins),  and  $2 < h < 4\,$ 
(single spin).}
\end{figure}

\begin{figure}
\vskip 2cm
\hskip -0.5cm
{\centering \resizebox*{4.5in}{!}{\includegraphics*{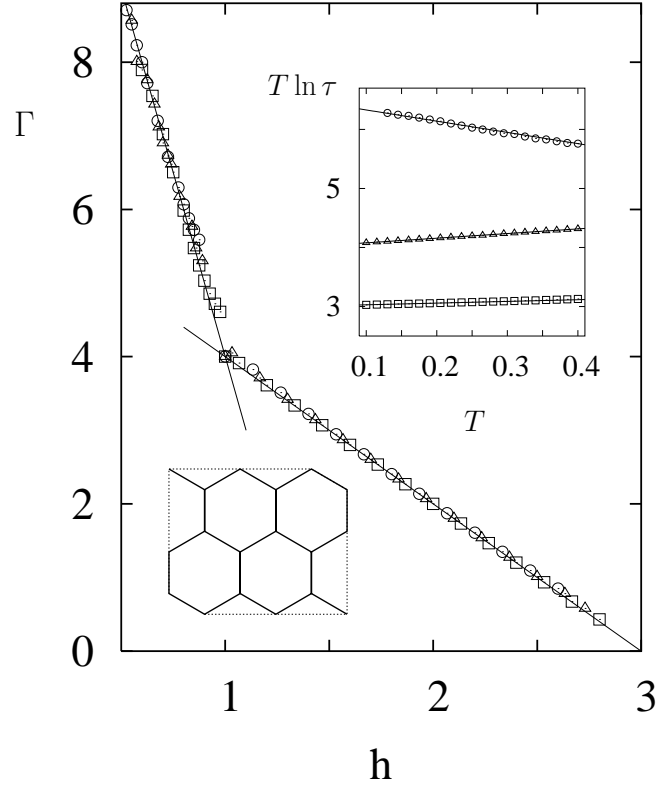}}}
\vskip -2cm
\caption{Relaxation parameters of honeycomb lattices estimated from 
the diagonalization of Eq.\,(\ref{SGlaub}) using the 18-spin cluster
depicted by the lower inset. As in Fig. 1, the amplitudes yielding the data 
collapse in the main panel [\,$T = 0.2\,$ (squares), 0.3 (triangles) and 0.4
 (circles)\,], were inferred from the slopes of the upper inset. The latter 
refer respectively to $h = 0.75\,$ (top), 1 (middle), 1.5 (bottom)
and are representative of the field regimes summarized in 
Eq.\,(\ref{hard-honeycomb}) (solid lines of main panel).}
\end{figure}

\begin{figure}
\vskip 2cm
\hskip -0.5cm
{\centering \resizebox*{4.55in}{!}{\includegraphics*{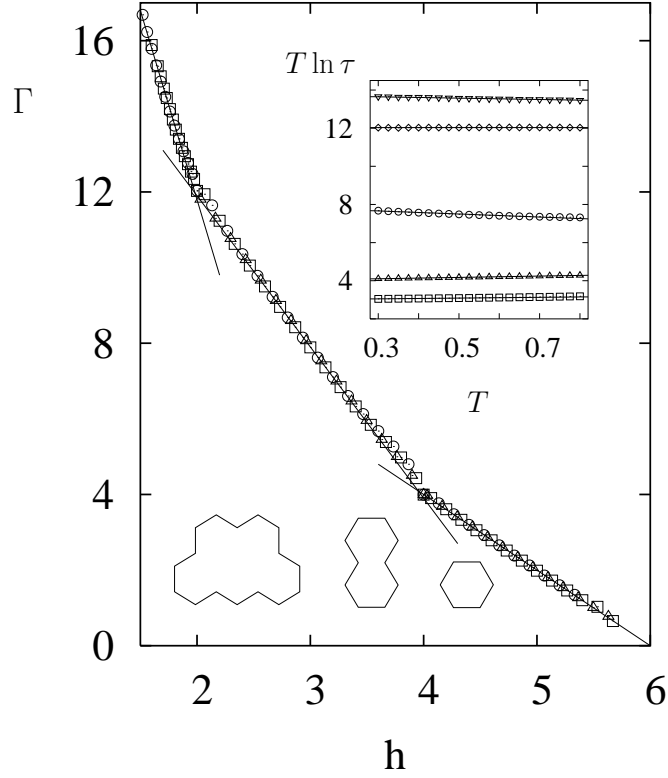}}}
\vskip -2cm
\caption{Relaxation parameters of triangular lattices arising from the gap 
of Eq.\,(\ref{SGlaub}) in $4 \times 4$ and $5 \times 4\,$ spin clusters. 
As before, the data collapse was obtained from the slopes of the upper
 inset.  The former refers to $[\,T = 0.3\,$ (triangles), 0.4 (squares) and 
0.5 (circles)\,], and follows closely the field regimes given in
 Eq.\,(\ref{hard-triangular}), denoted by solid lines. The amplitudes 
resulting from the inset slopes are characteristic of the regimes identified 
in Eq.\,(\ref{hard-triangular}). Here, they refer to  $h =\,$ 1.8, 2, 3, 4, and 
4.5, in descending order. The size and shape of critical droplets are 
schematized below.  From left to right  they refer to 5-spins 
($1.5 \alt h < 2\,$), 2-spins $( 2 < h < 4\,$) and a  single spin 
($ 4 < h < 6\,$).}
\end{figure}

\begin{figure}
\vskip -1.5cm
\hskip -1.8cm
{\centering \resizebox*{4.3in}{!}{\includegraphics*{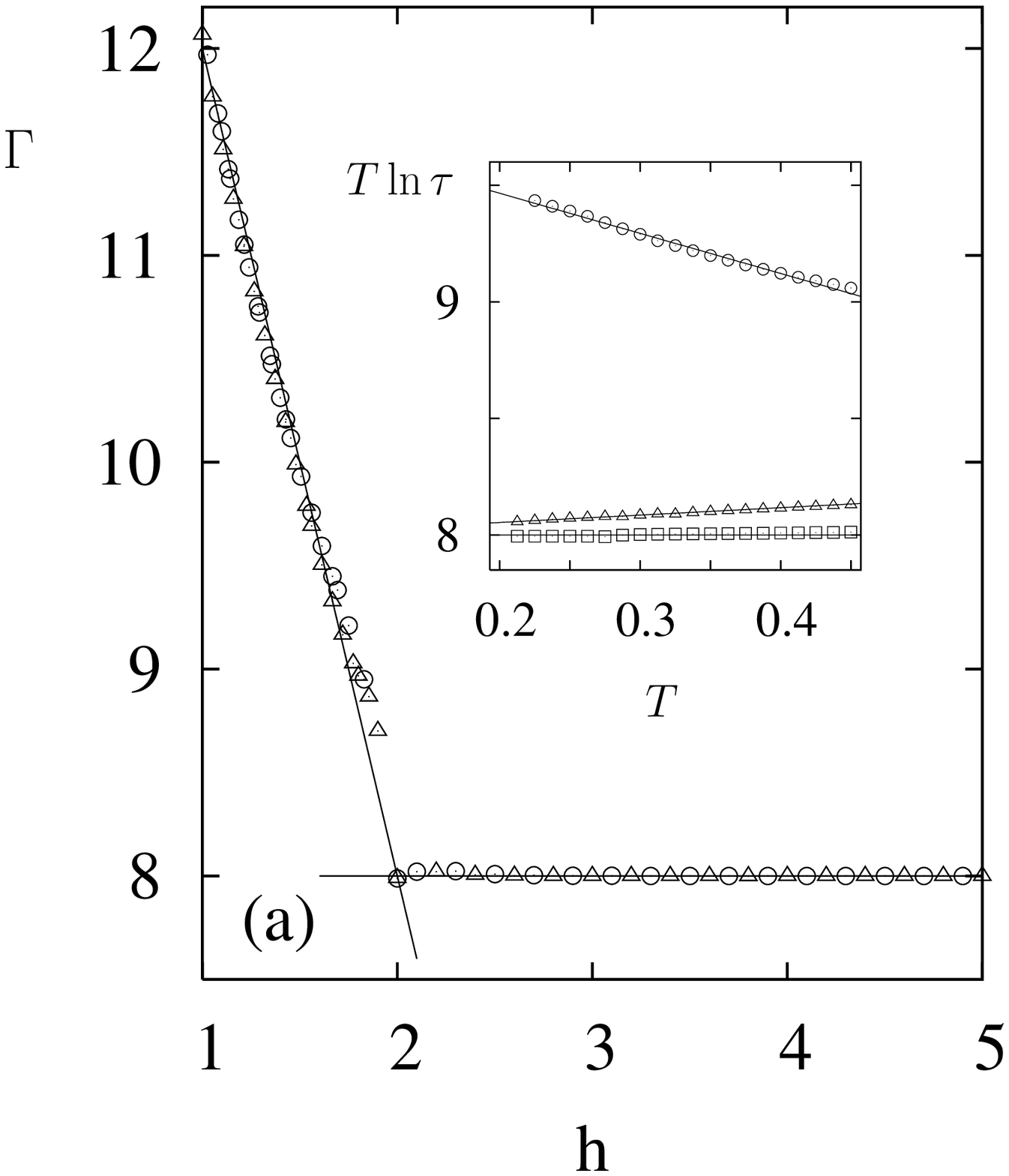}}}
\hskip -2.3cm
{\centering \resizebox*{4.3in}{!}{\includegraphics*{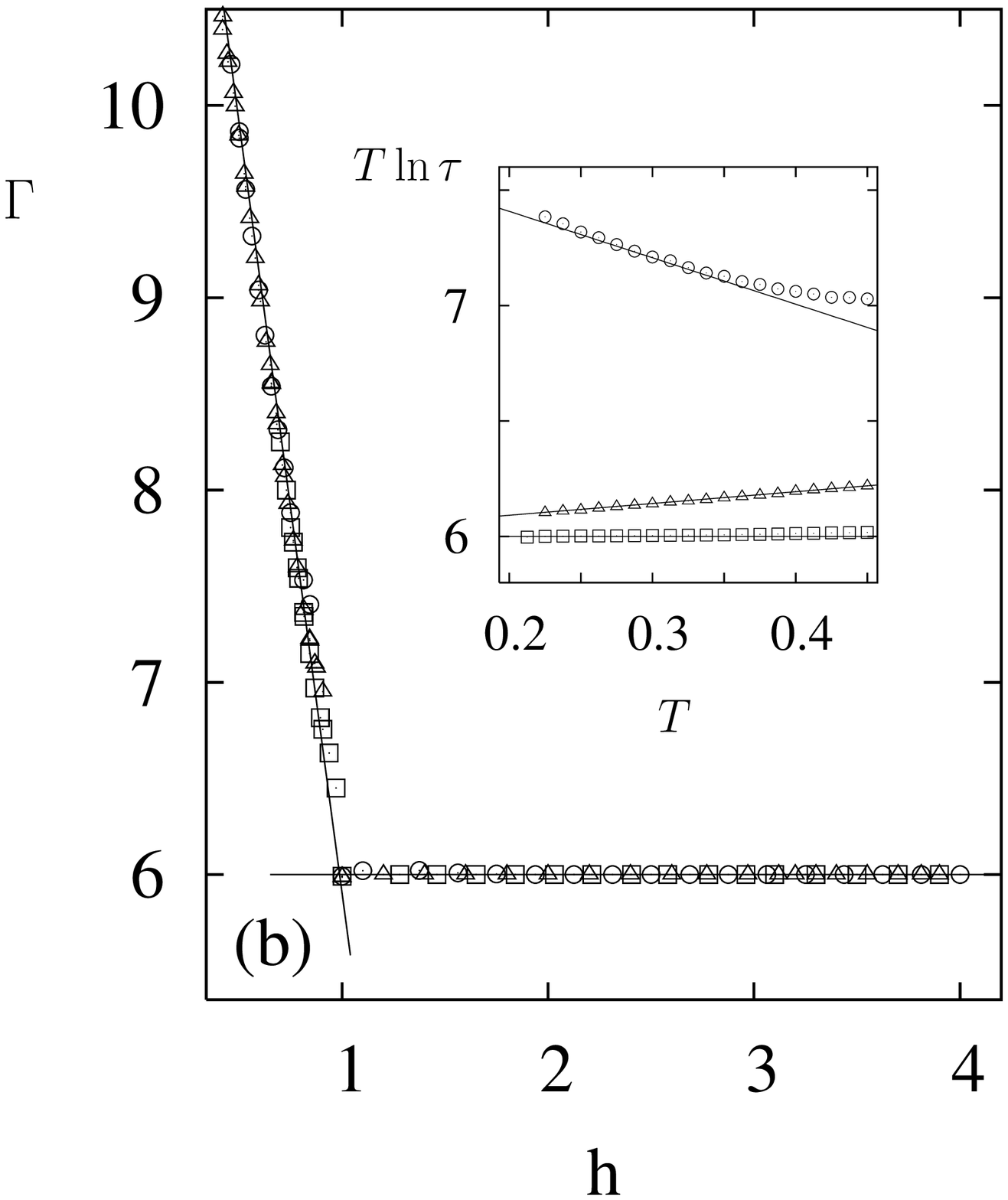}}}

\vskip -4.7cm
\hskip -10.5cm
{\centering \resizebox*{4.25in}{!}{\includegraphics*{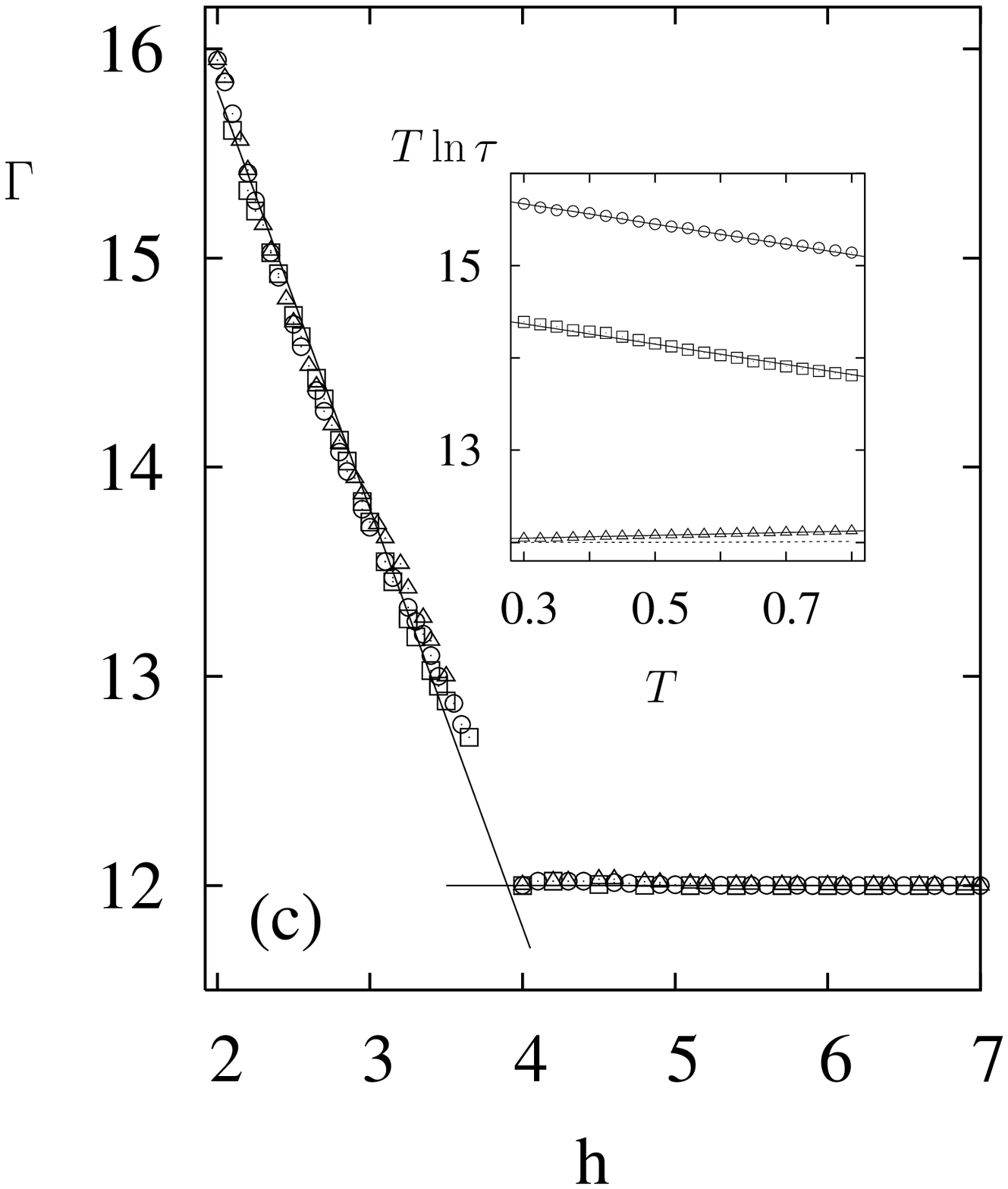}}}
\vskip -2cm
\caption{Relaxation parameters of (a) square, (b) honeycomb  and, 
(c) triangular lattices under the {\it modified} Glauber dynamics given in 
Eq.\,(\ref{rMG}). Solid lines in each case stand respectively for the 
regimes identified in Eqs.\,(\ref{soft-square}), (\ref{soft-honeycomb}) and, 
(\ref{soft-triangular}). In (a) and (b) they follow closely the data of 
$T = 0.4\,$ (circles), 0.3 (triangles) and, 0.2 (squares), obtained from 
the numerical diagonalization of Eq.\,(\ref{SMGlaub}). Similarly for (c), 
where  the points denote $T = 0.8\,$ (triangles), 0.6 (circles) and, 0.4 
(squares). In contrast to the standard  Glauber dynamic, here the
relaxation process remains active at large fields, i.e.  $\Gamma  > 0\,$ for
$h \gg J\,$.  As before, the amplitudes were derived from the slopes of 
the insets.  In descending order they refer respectively to (a) $h =1.5,\, 
2,\, 2.5\,$,  (b) $h=0.75,\, 1\,, 1.5\,$ and (c) $h = 2,\,2.5,\,4,\,5\,$ 
(horizontal line),  typical cases of  each situation.}
\end{figure}


\begin{thebibliography}{99}
\bibitem{factor} Certainly, for $h = 0\,$ both 
dynamics coincide except for a factor $2$, which is retained to 
follow the notation of Ref.\,[\onlinecite{Park}].
\end{thebibliography}
\end{document}